\documentclass[
    aip,
    amsmath,
    amssymb,
    reprint
]{revtex4-1}
\usepackage{amsmath,amssymb,amsfonts,mathrsfs}
\usepackage{amsthm}
\usepackage{graphicx}
\usepackage{dcolumn}
\usepackage{bm}
\usepackage{color}

\begin{document}
	
	\title{Factoring six-digit integers with superconducting quantum circuits}

    \author{Xi Zhang}
    \affiliation{National Laboratory of Solid State Microstructures, School of Physics, Nanjing University, Nanjing 210093, China}
    \affiliation{Shishan Laboratory, Suzhou Campus of Nanjing University, Suzhou 215000, China}
    \affiliation{Jiangsu Key Laboratory of Quantum Information Science and Technology, Nanjing University, Suzhou 215163, China}
    \affiliation{Hefei National Laboratory, Hefei 230088, China}
 
	\author{Xinsheng Tan}
    \email{tanxs@nju.edu.cn}
	\affiliation{National Laboratory of Solid State Microstructures, School of Physics, Nanjing University, Nanjing 210093, China}
    \affiliation{Shishan Laboratory, Suzhou Campus of Nanjing University, Suzhou 215000, China}
    \affiliation{Jiangsu Key Laboratory of Quantum Information Science and Technology, Nanjing University, Suzhou 215163, China}
    \affiliation{Synergetic Innovation Center of Quantum Information and Quantum Physics, University of Science and Technology of China, Hefei, Anhui 230026, China}
    \affiliation{Hefei National Laboratory, Hefei 230088, China}

	\author{Yang Yu}
	\email{yuyang@nju.edu.cn}
	\affiliation{National Laboratory of Solid State Microstructures, School of Physics, Nanjing University, Nanjing 210093, China}
    \affiliation{Shishan Laboratory, Suzhou Campus of Nanjing University, Suzhou 215000, China}
    \affiliation{Jiangsu Key Laboratory of Quantum Information Science and Technology, Nanjing University, Suzhou 215163, China}
    \affiliation{Synergetic Innovation Center of Quantum Information and Quantum Physics, University of Science and Technology of China, Hefei, Anhui 230026, China}
    \affiliation{Hefei National Laboratory, Hefei 230088, China}
	
	\begin{abstract}

        Integer factorization is a central computational problem with important applications in public-key cryptography. Here, we demonstrate a quantum factorization protocol using a superconducting circuit. Microwave drives are used to engineer a highly tunable effective two-level Hamiltonian whose eigenvalues can be measured spectroscopically. The target integer \(N\) and each candidate factor pair (\(p\),\(q\)) are encoded into the amplitudes, frequencies, and phases of the applied microwave fields. By scanning the candidate pairs while monitoring the spectral response at zero energy, we identify the factor pairs of integers up to six digits. The protocol requires neither two-qubit gates nor quantum entanglement. Its performance is currently limited by the precision of microwave control and the finite linewidth of the spectroscopic response. Improved control accuracy and longer coherence times would extend the accessible range of integers. 
	\end{abstract}

	\maketitle

	\bigskip

    Integer factorization is a problem in NP and underlies the widely used Rivest-Shamir-Adleman (RSA) public-key cryptosystem \cite{RSA}.
    Shor's algorithm shows that a quantum computer can factor integers exponentially faster than the best-known general-purpose classical algorithms \cite{shor1994algorithms}.
    This prospect has motivated substantial efforts to develop quantum computers based on a variety of physical platforms \cite{Divincenzo, Deutsch,Zoller_Ions,Kane_spin,makhlin,nakamura,vion}. The principle of Shor's algorithm has been demonstrated by the factorization of $N$ = 15 using Nuclear Magnetic Resonance (NMR) systems and $N=21$ using photonic systems \cite{N15,N21}. 
    However, many of these early experiments relied on prior knowledge of the factors or on problem-specific simplifications, limiting their applicability to general integer factorization.
    Moreover, the limited number and imperfect fidelity of currently available physical qubits make practical implementations of Shor's algorithm extremely challenging.
    
	An alternative approach to quantum factorization is adiabatic quantum computation, which is faster than classical computation for
	solving certain optimization problems \cite{farhi_ab,Mizel_ab}. 
    Several algorithms have been proposed \cite{Peng_ab,bocharov_ab,kebiao_ab,Schaller_ab}, increasing the factored number $N$ from 143 \cite{N143} to 56153 \cite{N56153}. Nevertheless, the prime factorization of an arbitrary number remains unrealized.	
    
    Here, we propose and experimentally demonstrate a factorization protocol implemented with a microwave-driven superconducting circuit. The applied microwave fields generate a highly tunable effective two-level Hamiltonian, and we identify candidate factor pairs by measuring its energy spectrum. 
    In contrast to experimental implementations of Shor’s algorithm, our protocol does not require two-qubit gates or quantum entanglement. 
    It therefore avoids several multi-qubit control requirements that pose major challenges for current quantum processors.
    
	The protocol is based on the diagonalization of a real symmetric $%
	2\times 2$ matrix. 
    Let \(N\) denote the integer to be factored, and let \(p\) and \(q\) be positive integers treated as candidate factors.
	We construct the matrix 
	\begin{equation}
	H=%
	\begin{pmatrix}
	p & \sqrt{N} \\
	\sqrt{N} & q \\
	\end{pmatrix}%
	,  \label{Ham}
	\end{equation}%
    The eigenvalues $x$ of $H$ are obtained by solving the
	characteristic equation:
	\begin{equation}
	\left\vert
	\begin{array}{cc}
	p-x & \sqrt{N} \\
	\sqrt{N} & q-x%
	\end{array}%
	\right\vert =0.  \label{eigen1}
	\end{equation}%
	Expanding the determinant in Eq.~\eqref{eigen1} gives:
	\begin{equation}
	(p-x)(q-x)-N=x^{2}-(p+q)x+pq-N=0.  \label{eigen2}
	\end{equation}%
    The two solutions of this equation are the eigenvalues of \(H\).
    The matrix \(H\) has a zero eigenvalue if and only if \(pq-N=0\), or equivalently \(N=pq\).
    The factorization problem can therefore be formulated as a search for a candidate pair \((p,q)\) that yields a zero eigenvalue. 
    If the spectrum of \(H\) can be measured directly, candidate pairs can be scanned until the zero-eigenvalue condition is satisfied.
    
    Unfortunately, on a classical computer, solving the characteristic equation while scanning $p$ and $q$ may take longer than testing candidate pairs directly by trial division.
    By contrast, a superconducting qubit physically realizes a two-level Hamiltonian whose eigenvalues are reflected in its measurable energy spectrum, providing a direct experimental means of evaluating the zero-eigenvalue condition.

\begin{figure}[tbph]
		\includegraphics[width=\columnwidth]{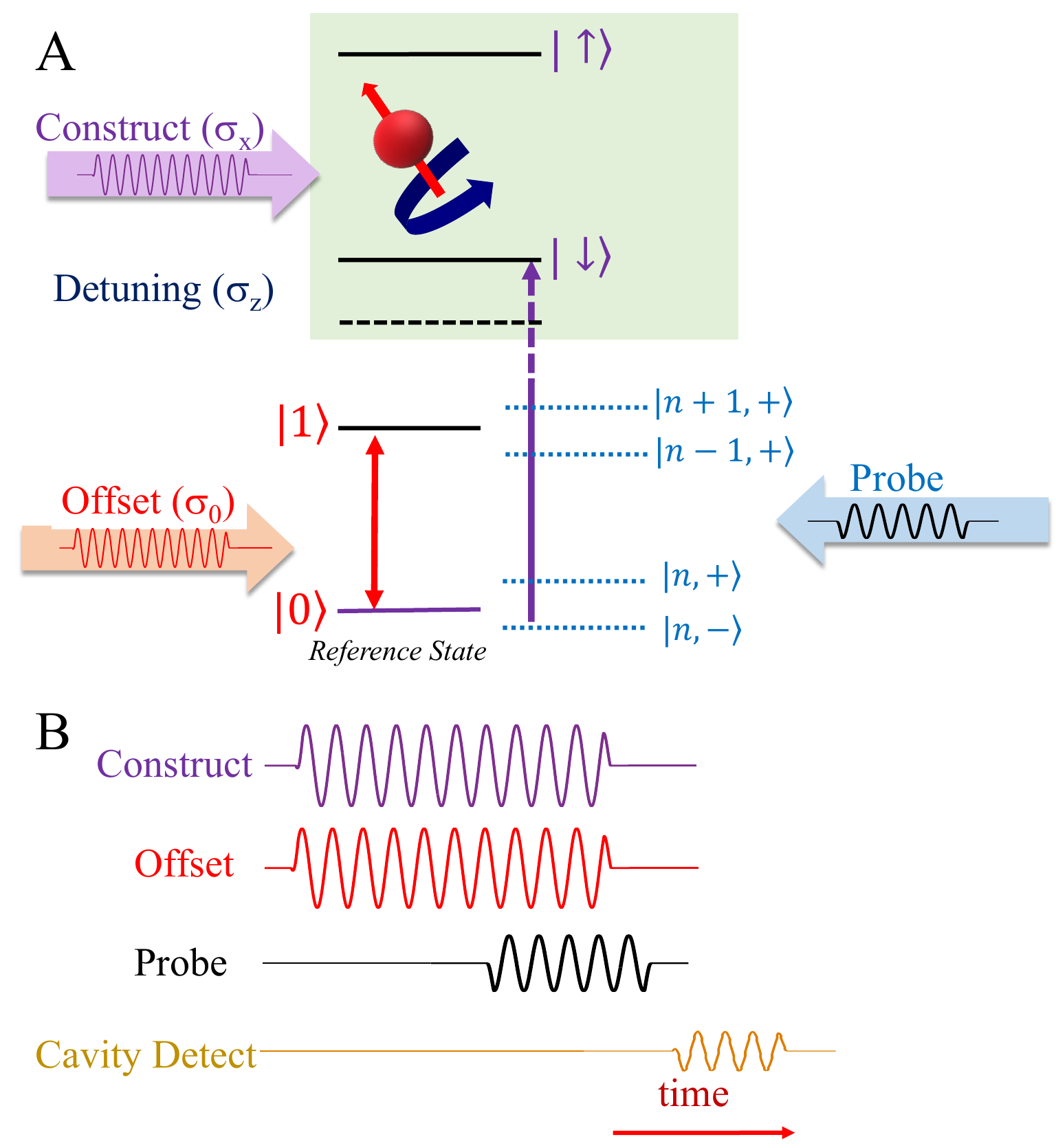}
		\caption{\textbf{(A)} Schematic of the energy-level structure of quantum prime factorization using superconducting circuits. The four energy levels are grouped into two subspaces and labeled $\{|0\rangle, |1\rangle \}$, $\{|\uparrow\rangle, |\downarrow\rangle \}$. The latter is treated as a spin-1/2 artificial atom.  
        The construct drive controls the transverse coupling \(B_1\) and the detuning term \(B_3\), whereas the offset drive shifts the reference energy associated with \(B_0\).
        The factors $p$ and $q$ are determined by spectroscopic measurement.  
        \textbf{(B)} Time profiles of the applied microwave pulses. The construct and offset pulses each last \(90~\mu\mathrm{s}\), whereas the probe pulse lasts \(50~\mu\mathrm{s}\). The qubit state is measured from the amplitude of the cavity-transmission signal.
		}
		\label{demon}
	\end{figure}
    
    To describe this physical implementation, we model the superconducting qubit as an effective two-level quantum system \cite{Yu-Science02,Lupascu-PRL2015}. In the presence of an effective magnetic field, its Hamiltonian can be written as ($\hbar =1$ for simplicity)
	\begin{equation}
	\hat{H}=\sum_{i=0}^{3}B_{i}\sigma_{i}/2,\label{Ham1}
	\end{equation}
    where $\sigma_{0}$ is the identity matrix and $\sigma_{1,2,3}$ are the three Pauli matrices.
    In conventional descriptions of qubit dynamics, $B_{0}$ is usually set to zero because it contributes only an overall energy offset. 
    In our protocol, however, we tune it by shifting the zero-energy point because $p$ and $q$ depend on the $B_{0}$ term.
    The coefficients $B_{1}$ and $B_{2}$ are proportional to the amplitude of an RF (transverse) magnetic field whose frequency is resonant with the energy-level spacing.
    The coefficient $B_{3}$ is proportional to a DC (longitudinal) magnetic field that defines the energy-level spacing of the two-level system.  
    
    To simplify the experimental implementation and improve the control accuracy, we realized these effective fields using microwave drives rather than static magnetic fields.
    In the rotating frame, a microwave-driven two-level system is described by a Hamiltonian that takes the same form as Eq.~\eqref{Ham1}.
    The transverse coefficients \(B_1=\Omega_1\) and \(B_2=\Omega_2\) are determined by the in-phase and quadrature components of the microwave drive, which control rotations about the \(x\) and \(y\) axes of the Bloch sphere, respectively. 
    These coefficients can be continuously tuned by varying the amplitude and phase of the microwave drive.
    The longitudinal coefficient $B_{3}=\Delta$ is set by the detuning between the microwave frequency and the transition frequency of the two-level system.
    
	We implemented the Hamiltonian in Eq.~\eqref{Ham} using a three-dimensional (3D) superconducting transmon system \cite{paik_3d,Tan_Maxwell}. The transmon, which consisted of a single Josephson junction and two pads (250 $\mu $m $\times
	$ 500 $\mu $m), was mounted in a rectangular aluminum cavity whose $TE_{101}$ mode had a resonance frequency of \(9.026~\mathrm{GHz}\).
    In our experiments, the cavity was used primarily to control and read out the transmon, and the system was designed to operate in the dispersive regime. 
    The entire device assembly was cooled to a base temperature of 30 mK in a dilution refrigerator. 
    The system dynamics were modeled within the framework of circuit quantum electrodynamics (circuit QED), which describes the interaction between an artificial atom and microwave fields\cite{Nori,blais_qed,wallraff,You_qed}.
    Microwave fields were used to control the quantum states of the transmon.
    In-phase and quadrature (IQ) mixers combined with a 1 GSa/s arbitrary waveform generator (AWG) were used to modulate the amplitudes, frequencies, and phases of the microwave pulses~\cite{Tan_pt}. 
    The measurement was performed using a \textquotedblleft high-power readout" scheme \cite{Tan_Maxwell}. 
    When a strong microwave tone resonant with the cavity was applied, the amplitude of the transmitted signal depended on the transmon state because of the nonlinear response of the cavity-QED system.

\begin{figure}[tbph]
		\includegraphics[width=\columnwidth]{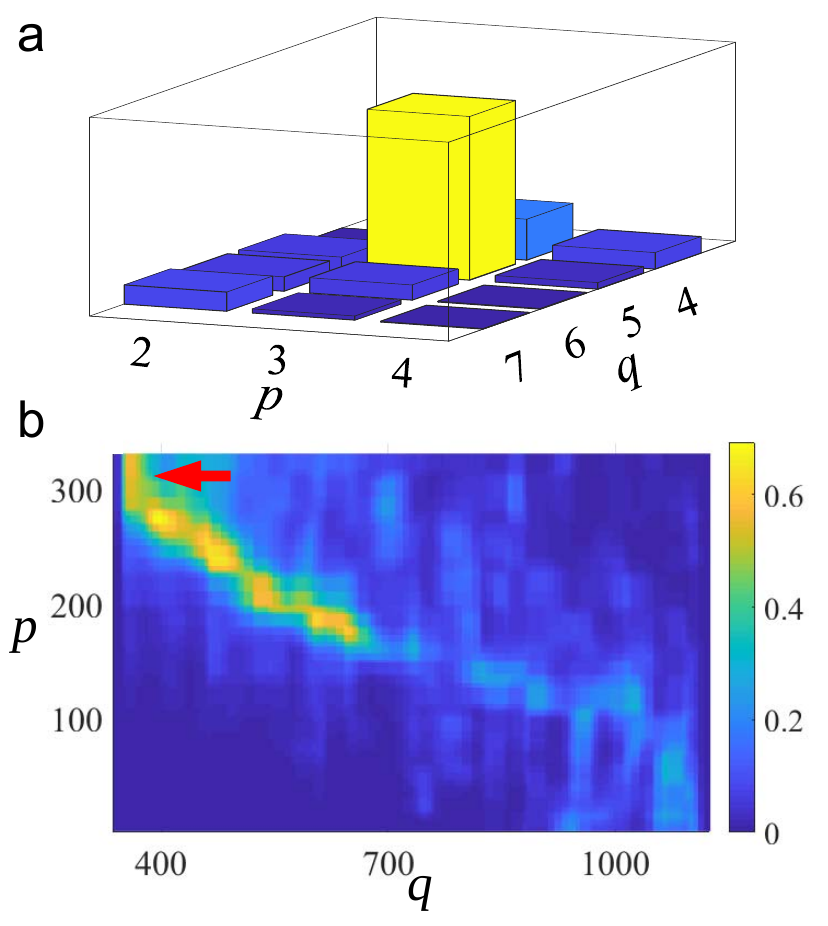}
		\caption{\textbf{(A)} Experimental result for the prime factorization of $N$ = 15. 
        The bar height represents the result (population of the excited state) of the spectroscopic measurement. 
        The yellow bar shows that the factor pair $(3,5)$ was obtained with relatively high probability by our algorithm.
			\textbf{(B)} Experimental result for the prime factorization of $N$ = 113569. The population of the excited state at the zero-energy point is plotted as a function of $(p,q)$. 
            Classical verification identified the correct factor pair $(337,337)$, marked by the red arrow.
		}
		\label{N15}
	\end{figure}
    
	Conventionally, a transmon coupled to a cavity exhibits a weakly anharmonic multilevel energy spectrum. In our experiments, we considered the four lowest bare energy levels of the transmon, $|0\rangle$, $|1\rangle$, $|2\rangle$, and $|3\rangle $, which are shown as black solid lines in Fig.~\ref{demon}A. The corresponding transition frequencies between adjacent energy levels were $\omega _{01}/2\pi =$ 7.1671 GHz, $\omega _{12}/2\pi =$ 6.8310 GHz and $\omega _{23}/2\pi =$ 6.340 GHz, respectively. These frequencies were independently calibrated using saturation spectroscopy.
    The corresponding energy relaxation times of the transmon were $T_{1}^{01}\sim $ 15 $\mu s$, $T_{1}^{12}\sim $ 12 $\mu s$ and $T_{1}^{23}\sim $ 10 $\mu s$, while the dephasing times were $T_{2}^{\ast 01}\sim $ 6.3 $\mu s$, $T_{2}^{\ast 12}\sim $ 5.5 $\mu s$ and $T_{2}^{\ast 23}\sim $ 4.0 $\mu s$. 
    
    To realize the Hamiltonian in Eq.~\eqref{Ham1}, we divided the four bare states into two subspaces $\{|0\rangle, |1\rangle \}$ and $\{|2\rangle, |3\rangle \}$, relabeling the latter as $\{|\uparrow\rangle, |\downarrow\rangle \}$. 
     Microwave control of the $\{|0\rangle, |1\rangle \}$ subspace was used to tune the $B_0$ term. In our experiments, two microwave drives, denoted as the ``construct'' and ``offset'' as shown in Fig.~\ref{demon}, were applied to the transmon. These drives coupled to the $\{|\uparrow\rangle, |\downarrow\rangle\}$ and $\{|0\rangle$, $|1\rangle\}$ subspaces, respectively. 
    Each drive produced microwave-dressed eigenstates within the corresponding subspace~\cite{Tan_pt}.
    The eigenstates in the $\{|\uparrow\rangle, |\downarrow\rangle\}$ subspace were selected as the two levels of an effective spin-1/2 system.  
    Therefore, the $B_1$ and $B_3$ terms in Eq.~\eqref{Ham1}, which are related to the amplitude and frequency of the construct drive~\cite{Tan_pt,Tan_Maxwell}, are tunable within the $\{|\uparrow\rangle, |\downarrow\rangle\}$ subspace.  
    
    Meanwhile, the dressed state $%
	\left\vert n,-\right\rangle $ in the  $\{|0\rangle$, $|1\rangle\}$ subspace was chosen as the reference level. 
    This means the effective zero-energy of the artificial atom, which is associated with the $B_0$ term, can be manipulated by shifting the reference level $\left\vert n,-\right\rangle $. 
    As a function of the applied microwave drives, the energy separation between the dressed reference state \(|n,-\rangle\) and the bare state \(|2\rangle\), which sets the energy reference for the effective spin-\(1/2\) subspace, is given by
	\begin{equation}
	E=\omega_{01}+\omega_{12}+\frac{\Delta_{01}}{2}+\frac{\sqrt{\Delta_{01}^2+\Omega_{01}^2}}{2},
	\end{equation}
	where $\omega_{01}$($\omega_{12}$) is the energy spacing between states $|0\rangle$($|1\rangle$) and $|1\rangle$($|2\rangle$); $\Delta_{01}=\omega_d-\omega_{01}$ is the detuning between the applied microwave frequency and the energy spacing; and $\Omega_{01}$ is the Rabi frequency induced by the applied microwave drive. 
    The effective Hamiltonian in Eq.~\eqref{Ham1} could thus be engineered by calibrating the frequencies, amplitudes, and phases of the applied microwave fields.
    
    We directly measured the eigenenergies of the Hamiltonian in Eq.~\eqref{Ham1} by applying an additional probe microwave and performing a spectroscopic measurement. 
    The pulse sequence is shown in Fig.~\ref{demon}B.
    The construct and offset drives were each applied for \(90~\mu\mathrm{s}\), while the probe pulse had a duration of \(50~\mu\mathrm{s}\).
    We then inferred the transmon state from the cavity transmission amplitude.
    The positions of the resonant absorption peaks in the resulting spectrum yielded the eigenenergies of the effective Hamiltonian. 
    According to Eq.~\eqref{eigen2}, one eigenenergy is zero if and only if $N=p\times q$. Therefore, we only need to monitor the resonant peak (or the excited-state population) at the zero-energy reference in our protocol.
    
    The practical procedure for factoring a given integer \(N\) is as follows:
    
    1) Encode \(\sqrt{N}\) in the real off-diagonal matrix element by setting the transverse coefficient \(B_1\).
	
    2) Encode a candidate pair \((p,q)\) in the diagonal matrix elements through the coefficients \(B_0\) and \(B_3\).
	
    3) Measure the excited-state population at the zero-energy reference point. If a resonance peak is observed, record \((p,q)\) as a candidate factor pair and verify whether \(pq=N\).
	
    4) If no peak is observed, select the next candidate pair and repeat the measurement until the factors are identified.
    
    For a given \(N\), we swept $p$ and $q$ over the range $ (2, \lceil \sqrt{N} \rceil ) \times (\lfloor \sqrt{N}  \rfloor, \lceil N/2 \rceil )$ in our experiments.
	 By comparing Eq.~\eqref{Ham} with Eq.~\eqref{Ham1}, we find that $\{p,q,\sqrt{N}\}$ correspond to $\{(B_0+B_3)/2, (B_0-B_3)/2,B_1/2\}$. For an extremely large $N$, it is convenient to normalize the Hamiltonian by $\sqrt{N}$: 
	
	\begin{equation}
	H=\Omega_0
	\begin{pmatrix}
	\frac{p}{\sqrt{N}} & 1 \\
	1 &  \frac{q}{\sqrt{N}} \\
	\end{pmatrix}%
	,  \label{Ham2}
	\end{equation}
	where $\Omega_{0}$ is the reference frequency scale, and we set \(\Omega_0/2\pi=5~\mathrm{MHz}\) in the experiments.

	We have demonstrated our protocol using several examples. We first considered $N$ = 15.
    Figure~\ref{N15}A shows the measured excited-state population at zero energy for different candidate pairs \((p,q)\).
    The zero-energy reference was defined by \(\omega_{01}+\omega_{12}+\Delta/2\), and the candidate integers were scanned over \(p\in[2,4]\) and \(q\in[4,7]\). A pronounced population peak was observed at \((p,q)=(3,5)\), thereby correctly identifying (3,5) as the factor pair of 15. This result demonstrates the basic operating principle of the protocol without encoding the known factors directly into the experimental sequence.
    
    With our present experimental setup, we find that it is possible to factor six-digit integers. Figure~\ref{N15}B shows the experimental result for $N$ = 113569. 
     The excited-state population at zero energy exhibited multiple peaks.
    Then we checked the corresponding candidate pairs classically, yielding the correct factor pair \((p,q)=(337,337)\), as indicated by the red arrow.
    Although classical verification introduces additional overhead, spectroscopic screening restricts this verification to a subset of the candidate pairs.
    The integer factored here is more than three orders of magnitude larger than those factored in previous experimental demonstrations of Shor’s algorithm.
    
    In our current setup, two experimental limitations restrict the size of \(N\).
    The first is the limited precision of microwave control, which depends on the performance of the microwave-control hardware. 
    In our implementation, the discrete encoding step is $\Omega_{0}/(2\pi\sqrt{N})$, corresponding to approximately 15 kHz for $\Omega_{0}/2\pi$ = 5 MHz and $N$ = 113569. 
    The standard deviation of the microwave amplitude was about \(1\%\) of \(\Omega_0\), corresponding to an effective uncertainty of approximately \(50~\mathrm{kHz}\). Because this uncertainty exceeds the encoding step, fluctuations in the microwave control broaden the distribution of candidate responses.
    
    The second limitation is the linewidth of the resonant peaks, which is related to the coherence time of the transmon qubit by $2 \pi \delta_{HWHM}={T_2}'^{-1}=\sqrt{T_2^{-2}+\Omega_{Rabi}^2 T_1/T_2}$ ~\cite{Schuster}. 
    In our experiments, $\Omega_{Rabi}$ was small, resulting in a resonance linewidth of approximately $T_2^{-1}$. 
    Given the measured transmon coherence time of approximately 5 $\mu s$, the corresponding HWHM linewidth was about \(30~\mathrm{kHz}\), exceeding the \(15~\mathrm{kHz}\) encoding step. 
    This broadening substantially increased the number of candidate factor pairs $(p,q)$ satisfying $N\approx pq$.
    
\begin{figure}[t]
		\includegraphics[width=\columnwidth]{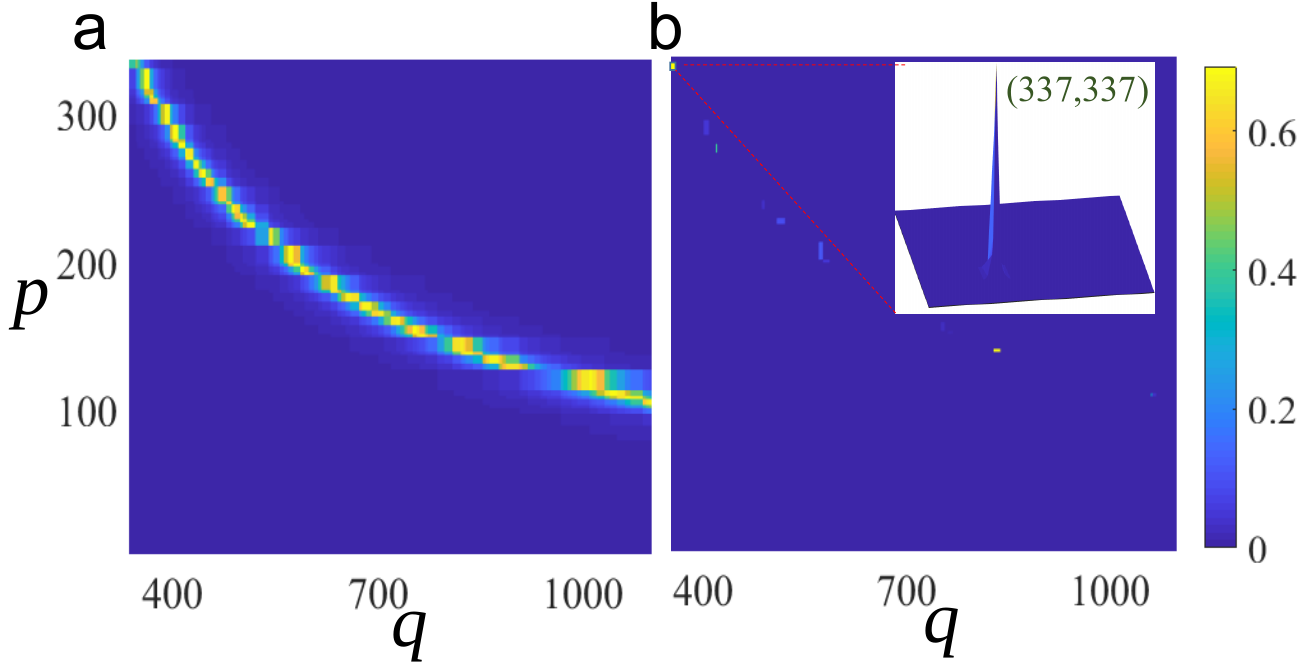}
		\caption{Numerical simulation for $N$ = 113569 with different microwave-drive strengths and coherence times. \textbf{(A)} $\Omega_{0}/2\pi$ = 5 MHz, $T_2$ = 5 $\mu s$, \textbf{(B)} $\Omega_0/2\pi$ = 50 MHz, $T_2$ = 100 $\mu s$. With longer coherence times and better microwave control, the candidate-factor distribution becomes more sharply localized.
		}
		\label{Ideal}
	\end{figure}

	Compared with Shor's algorithm, our method requires only single-qubit operations instead of two-qubit gates and quantum entanglement. 
    Because implementing high-fidelity multi-qubit control within finite coherence times remains experimentally challenging,  our method may be more readily implemented on current quantum hardware, as it relies primarily on single-qubit control and readout.
    To evaluate the potential improvement in the efficiency of the protocol, we performed numerical simulations assuming improved quantum-control accuracy and reduced decoherence.
    As shown in Fig.~\ref{Ideal}, we considered two parameter sets: \(\Omega_0/2\pi=5~\mathrm{MHz}\) and \(T_2=5~\mu\mathrm{s}\) for the first set, and \(\Omega_0/2\pi=50~\mathrm{MHz}\) and \(T_2=100~\mu\mathrm{s}\) for the second.
    The first parameter set closely matches the values used in our experiment, whereas the second set consists of optimized parameter values that can be achieved by improving the measurement circuits and fabrication process~\cite{Rigetti}. 
    Figure~\ref{Ideal} shows that a larger $\Omega_{0}$ and a longer coherence time $T_2$ produce a response concentrated around substantially fewer candidate factor pairs.
    The resulting reduction in the number of candidate pairs requiring verification corresponds to an improvement of approximately two orders of magnitude in the efficiency of the protocol.

    A potential advantage of our protocol over conventional classical procedures is that candidate factor pairs are evaluated through direct Hamiltonian encoding and spectroscopic readout, without explicitly solving the characteristic equation for each pair. Specifically, $p$ and $q$ are scanned while the excited-state population at the zero-energy reference is monitored. This signal-monitoring procedure can, in principle, be performed rapidly. Moreover, multiple qubits could be operated in parallel to scan different regions of the candidate space, potentially reducing the overall acquisition time. Nevertheless, the computational scaling of the protocol and any practical advantage over classical factorization algorithms remain to be established. Further improvements in scalability, control fidelity, and coherence would be required before the implications of this approach for cryptographic applications could be assessed.

    This work was supported by the National Key R\&D Program of China (Grant Nos. 2022YFA1405304 and 2024YFA1409300), the National Natural Science Foundation of China (Grant Nos. 12504588, U21A20436, and 12074179), the National Science and Technology Major Project for Quantum Science and Technology (Grant No. 2021ZD0301702), the Natural Science Foundation of Jiangsu Province (Grant Nos. BE2021015-1, BK20232002, and BK20233001), and the Natural Science Foundation of Shandong Province (Grant No. ZR2023LZH002).

    \section*{AUTHOR DECLARATIONS}

    \subsection*{Conflict of Interest}

    The authors have no conflicts to disclose.

    \subsection*{Author Contributions}

    Xi Zhang:  Data curation; Formal analysis;  Writing – original draft (equal); Writing – review \& editing (equal).

    Xinsheng Tan: Conceptualization; Methodology; Investigation;  Supervision (lead); Writing – original draft (lead); Writing – review \& editing (lead).

    Yang Yu: Supervision (equal); Resources; Writing – review \& editing (equal).
    

\end{document}